# Leveraging Multiple CNNs for Triaging Medical Workflow

Lakshmi A. Ghantasala
*Electrical and Computer Engineering*
*Purdue University*
West Lafayette, USA
lghantas@purdue.edu

ABSTRACT

High hospitalization rates due to the global spread of Covid-19 bring about a need for improvements to classical triaging workflows. To this end, convolutional neural networks (CNNs) can effectively differentiate critical from non-critical images so that critical cases may be addressed quickly, so long as there exists some representative image for the illness. Presented is a conglomerate neural network system consisting of multiple VGG16 CNNs; the system trains on weighted skin disease images re-labelled as 'critical' or 'non-critical', to then attach to input images a critical index between 0 and 10. A critical index offers a more comprehensive rating system compared to binary critical/non-critical labels. Results for batches of input images run through the trained network are promising. A batch is shown being re-ordered by the proposed architecture from most critical to least critical roughly accurately.

I. INTRODUCTION

The global spread of Covid-19 has had healthcare systems over-run with patients requiring hospitalization. In cases of crises, it becomes paramount that there exists some efficient method which may quicken the time a critical-condition patient spends awaiting treatment. In this vein, neural networks may be trained to distinguish critical cases from non-critical cases, rather than as diagnosis machines as they have traditionally been used [1]. Where a professional must spend precious time to determine the severity of a batch of cases, networks could quickly order such lists from most to least critical. This not only leaves diagnosis to the professional, but the lower accuracy requirement of the task gives networks leeway in case of error. The worst-case scenario for mis-triaging is a random ordering of cases, which may be reviewed by a professional regardless. For triaging via neural network to be effective against an illness like covid-19, covid-19 must be detectable in some form of image. With no such image dataset available for viruses, the dataset used was a skin lesion dataset, where RGB images suffice to be able to characterize the illness.

This paper proposes a conglomerate neural network system consisting of 5 separate VGG16 [2] CNNs such that multiple classifications may be taken into account to identify the criticality of skin lesions from RGB images. A publicly available skin disease dataset was used, though the ideas presented here can easily be applied to characteristic image inputs for different illnesses. Each of these networks was pre-trained on the ImageNet database [3][4], and has been partially fine-tuned with binary-labelled (critical and non-critical) inputs, similar to those used in previous efforts [5]. The base VGG16 network will be explained in Section II.A. The structure of the conglomerate network, and the use of multiple classifications will be explained in section II.B. The HAM10000 [6] skin disease dataset was used to train these models. The nature of this dataset, as well as modifications made to accommodate a critical vs non-critical use case will be outlined in section II.C. Finally, results and discussion will be Section III.

## II. CONGLOMERATE NEURAL NETWORK

### A. VGG16 Convolutional Neural Network

Among the highest performing CNNs in the recent past is the VGG16 [2] network architecture. This network can be split into 8 blocks, labelled in Figure 1. The network alternates between convolutional layers and max pooling layers to gradually shrink the input dataspace into a feature-space which consists of larger, composite pixel features in the data. Max pooling layers reduce the xy-size of the data space, while convolutional layers using the ReLu activation function (1) dilute the feature space.

$$R(x) = \max(0, x) \qquad (1)$$

Capping these alternating convolutional and max pooling layers are 2 fully connected (fc) layers and a softmax layer to output a probability of the input being of some class.

VGG16 networks pre-trained on ImageNet were used, as the feature set would be much more diverse than one learned by training solely on a limited medical dataset. Two modifications were made to the pretrained architecture. The fully connected layers were replaced with untrained layers of the same dimension, and the capping 1x1x1000 softmax layer was replaced with a softmax layer of dimension 1x1x2, sufficient for the critical/non-critical classification requirement. Training on the networks was conducted by freezing blocks 1-

3 while allowing weights in blocks 4, 5 and the capping fully connected layers to be trainable. The architecture and training scheme are delineated in figure 1. Variations on this freeze scheme proved less effective than that presented here.

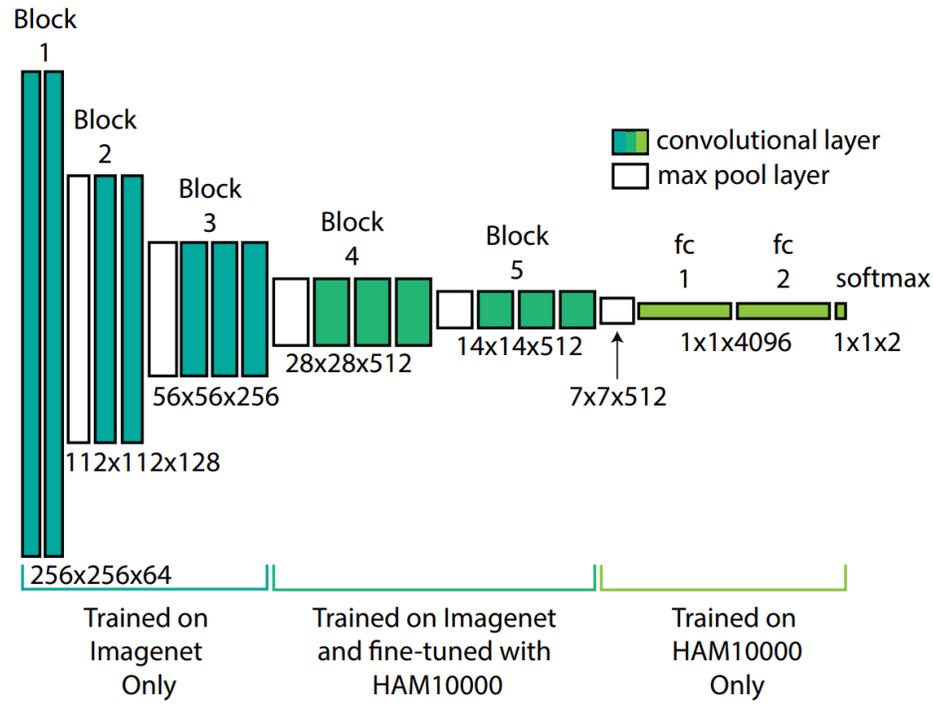

Fig. 1. Transfer learning was employed to make use of VGG16 CNNs pre-trained on the ImageNet database. Only the weights in the first three blocks were kept constant, while the rest of the network was allowed to train on the HAM10000 dataset. As the fully connected and softmax layers were zero-initialized, they do not contain any information from the image-net dataset.

A learning rate of 1e-6 was used. An NVIDIA GTX 1070 GPU with 8GB of v-ram was used. This greatly expedited the stochastic gradient descent algorithm used to train the network. Training and testing was carried out in python using the Keras API.

B. *Conglomerate Network Architecture*

Past designs for triaging radiology workflow are based on a single neural network [5]. When training on binary-labeled data, the output of these networks is most reliable when distinguishing between the two classes. For example, softmax outputs of 0.6 and 0.7 may be different due to noise, but both map to an image being classified as $> 0.5$. If the goal is to classify an image into two classes, the traditional method accurately classifies the images in both cases. However, this softmax output is less reliable when classifying *how much* the input image is of a class. While triaging workflow, this translates to reduced accuracy in determining which cases are more or less critical than others.

To ameliorate this issue, multiple neural networks' outputs [7] are averaged to obtain more accurate labels for input images. Variations are introduced between networks by modifying the class weights assigned to critical and non-critical images during training. Class weights are traditionally used to offset size mismatch in the training set between classes, to prevent a network from learning to classify a more abundant class at the cost of another. Here, this is the desired effect. Class weights are used to *introduce* a preference for networks to classify one class over another. The idea is to use class weights to bolster classification of critical cases in the first network, then slowly shift the class weights such that they bolster classification of non-critical images in the final network. Depending on how critical in image is, some networks will classify it as critical while others will classify it as non-critical. When averaging the classification of every network in the system, we obtain a critical index which may accurately represent how critical an image is.

This scheme retains the uncertainty leeway given to binary classification, while providing a higher resolution to how critical an image may be, rather than relying on the softmax output of a single network. This implementation uses 5 networks, assigning a critical index between 0 and 10 with intervals of 2 to input images. Figure 2 delineates the operation of this implementation.

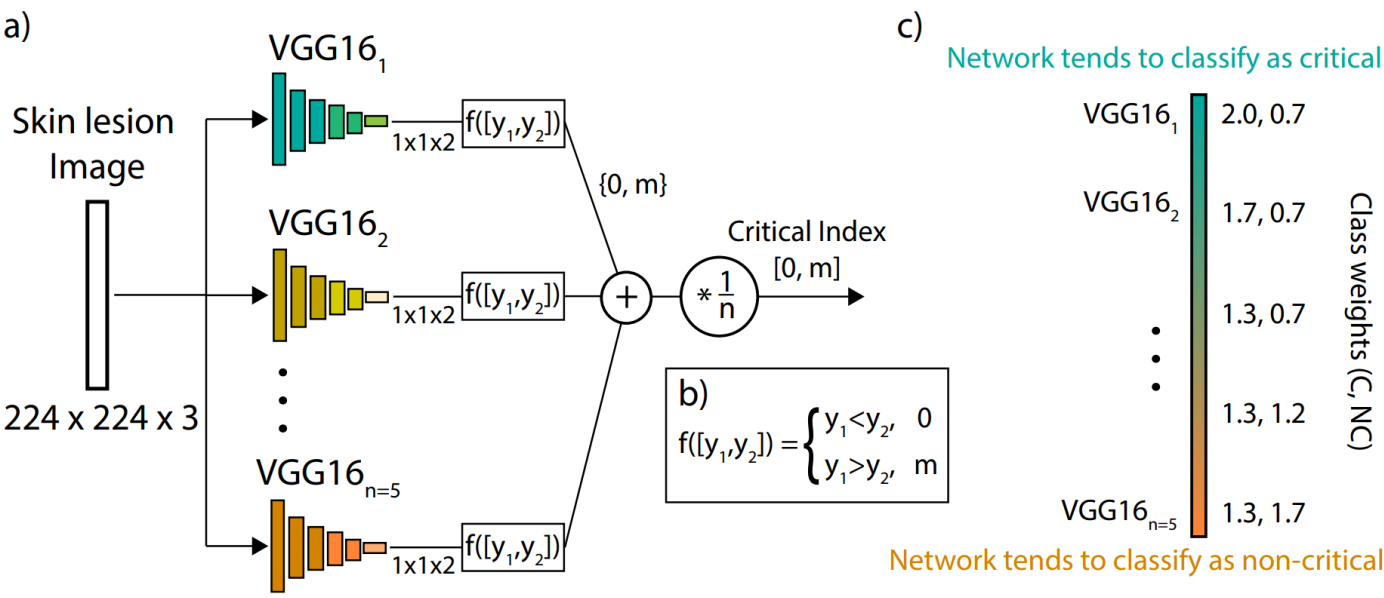

Fig. 2. (a) The conglomerate system makes use of 5 separately trained VGG16 networks, each trained on varying weighted data. A critical index is produced by averaging binary classifications by each network, set to m=10 following f([x,y]) in (b). (c) shows how the class weights are varied in this implementation. C is Critical and NC is Non-Critical.

The smallest delta $\Delta$ between consecutive critical index values for an *n*-network system is given by:

$$\Delta(n) = \frac{m}{n} \tag{2}$$

where m is the maximum critical index and n is the number of individual CNNs in the system. The scale implemented here uses m = 10. The more networks there are, the more continuous the critical index becomes.

The class weights are not normalized to 1,1 in figure 2. (c) as there is an inherent offset caused by unevenly sized classes. The 67% to 33% non-critical to critical ratio in the HAM10000 dataset, shown in figure 3, means to achieve unbiased training, the critical class will be weighted 1.3 and non-critical classes weighted 0.73.

*C. HAM10000 Dataset*

A published skin disease database known as HAM10000 [6] was used to fine-tune the pre-trained networks. The ideas presented in this paper, though implemented on skin lesion images, are suitable for any form of medical imaging. This database consists of 10015 images that cover 7 different skin cancers, shown in figure 3. Most of these images have been confirmed histopathologically, while the rest have been labelled in some combination of expert follow-up examination, expert consensus, or confirmation by in-vivo confocal microscopy [6]. Melanocytic nevi, the only benign skin lesion in the dataset, is labelled non-critical. The remaining malignant lesions are pulled together under one group and re-labelled critical.

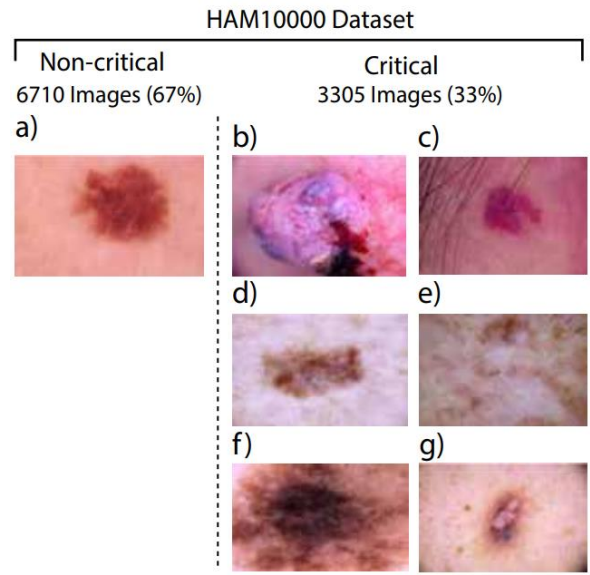

Fig. 3. The HAM10000 dataset has 10015 images of skin diseases in 7 categories that form a representative collection of diagnosable skin cancers [6] . The 7 categories are shown above: melanocytic nevi (a), benign keratosis-like lesions (b), vascular lesions (c), melanoma (d), Actinic keratoses and intraepithelial carcinoma / Bowen's disease (e), basal cell carcinoma (f), and dermatofibroma (g). Melanocytic nevi is a benign tumor, and has been reclassified as non-critical. The others have been classified as critical.

These images are packaged into 70 batches per epoch. Each batch is a random selection of 20 images, non-repeating within an epoch. Each network was trained for 20 epochs. A visible distinction between critical and non-critical lesions makes the HAM10000 dataset perfect for training a neural network to triage.

III. RESULTS AND DISCUSSION

To properly test this conglomerate network would require a dataset which has accurate labels on a scale of 0 to 10 for criticality. As such a dataset does not exist today, the network's efficacy was tested based on visual inspection of the critical index values assigned to images sequestered during training. Additionally, the conglomerate network was given a testing set of images to classify as critical or non-critical, as would traditionally be the case. A random set of 5 images with a critical label and a random set of 5 images with a non-critical label were tested, shown in figure 4. Please note that the first random selection was used, rather than re-running for optimal results. These 10 images mimic a real-life scenario in which a clinic must triage 10 cases.

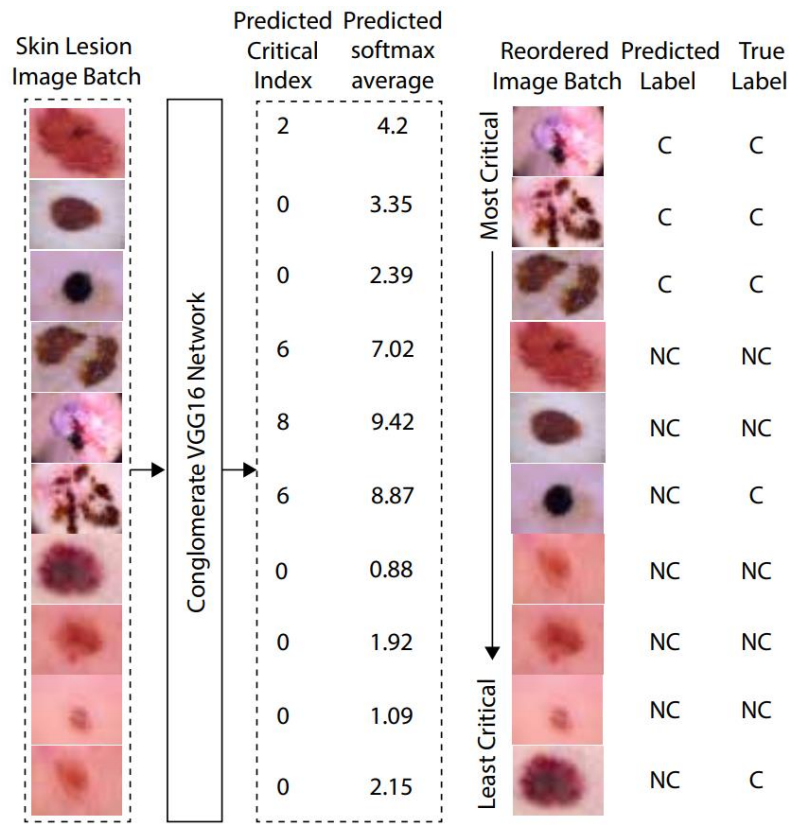

Fig. 4. The system roughly accurately sorts the input skin lesion image batch from most to least critical. The critical index of each image is shown, with ties broken by the average of the critical output node of each of the 5 networks. The 3 most critical cases are brought to the front, and are in proper order. C is Critical and NC is Non-Critical.

Figure 4 shows that the system performs as designed, assigning to the most critical images a higher critical index than non-critical images. The system even properly differentiates between criticality of critical images from a non-medical expert inspection, as the most severe case seems to have the highest critical index. The system tends to classify data as non-critical despite efforts to increase the class weight for critical data during training. To remedy this, a critical threshold of 3 was chosen, which splits the possible critical indices such that $\{0, 2\} \in$ NC and $\{4, 6, 8, 10\} \in$ C.

At a critical threshold of 3, the conglomerate network was tested on 1741 critical and 3274 non-critical images, resulting in an 82% accuracy in classifying critical images and a 69% accuracy in classifying non-critical images. It is important to remember that these metrics are a baseline, and the system goes one step further in determining how critical the image is.

## IV. CONCLUSION

Leveraging multiple CNNs may be a key step in advancing the accuracy of machine learning based triaging mechanisms, which is especially relevant in a time where hospitals are overrun with patients affected by Covid-19. Modifying class weights is one way to introduce variations between CNN's learning to predict criticality of cases. Different class weights encourage preferences for networks to classify one way over another. These different predictions can be averaged to produce a new metric, termed a critical index, which provides resolution into how critical a case may be. Results show that the conglomerate system is effective at distinguishing critical cases with 82% accuracy and can order images within the critical class effectively.

## V. ACKNOWLEDGEMENTS

I'd like to acknowledge Professor Jeffrey Siskind of Purdue University for motivating this effort.